\documentstyle[preprint,aps,epsf]{revtex}
\begin{document}

\draft \preprint{ARL Tech. Rep.}

\title{Transformation Properties of  the Lagrangian \\ and Eulerian Strain Tensors}
\author{Thomas B. Bahder}
\address{U. S. Army Research Laboratory \\
2800 Powder Mill Road \\
Adelphi, Maryland, USA  20783-1197}

\date{February 20, 2001}
\maketitle

\begin{abstract}
A coordinate independent derivation of the Eulerian and Lagrangian
strain tensors of finite deformation theory is given based on the
parallel propagator, the world function, and the displacement
vector field as a three-point tensor. The derivation explicitly
shows that the Eulerian and Lagrangian strain tensors are
two-point tensors, each a function of both the spatial and
material coordinates. The Eulerian strain is a two-point tensor
that transforms as a second rank tensor under transformation of
spatial coordinates and transforms as a scalar under
transformation of the material coordinates. The Lagrangian strain
is a two-point tensor that transforms as scalar under
transformation of spatial coordinates and transforms as a second
rank tensor under transformation of the material coordinates.
These transformation properties are needed when transforming the
strain tensors from one frame of reference to another moving
frame.
\end{abstract}

\section{Background}
The U.S. Army is developing an electromagnetic gun (EMG) for
battlefield applications.  During the past few years, on a
recurring basis, Dr. John Lyons (former ARL Director) and Dr. W.
C. McCorkle (Director of U. S. Army Aviation and Missile Command)
have requested that I look at some of the physics of the EMG.  In
the most recent request, I was asked to look at stresses in a
rotating cylinder. For the case of an elastic cylinder, this is a
classic problem that is solved in many texts on linear
elasticity~\cite{Love1944,LLelasticity1970,Nadai1950,Sechler1952,Timoshenko1970,VolterraGaines1971}.
However, when these derivations are examined closely, one finds
certain shortcomings~\cite{RotatingStrainReport}.  Therefore, I
spent some time looking at the problem of stresses in elastic
rotating cylinders. That work resulted in a
manuscript\cite{RotatingStrainReport}.  In the course of this
work~\cite{RotatingStrainReport}, I had to clearly understand the
transformation properties of the Largangian and Eulerian strain
tensors of finite deformation theory. I was quite dissatisfied
with the standard derivations of the Largangian and Eulerian
strain tensors because these derivations take either of two (both
unpalatable) approaches. In the first approach, shifter tensors
are used, which are often defined as inner products between two
basis vectors at {\it two different spatial
locations}~\cite{Eringen1962,Narasimhan1992}. In this approach,
basis vectors are not parallel transported to the same spatial
location before the inner product is carried out. This is
unpalatable, even in Euclidean space, unless one is using
Cartesian coordinates.   In the second approach, convected
(moving) coordinates are used, and vectors and tensors are
associated with a given {\it coordinate} in the convected (moving)
coordinate system, rather than being associated with a point in
the underlying space.

In the derivation that I present below, I avoid both of the
unpalatable features mentioned above.  I provide a coordinate
independent derivation of the Lagrangian and Eulerian strain
tensors based on standard concepts in differential geometry: the
parallel propagator, the world function, and the displacement
vector field as a three-point tensor.

The derivation that I present below is also useful for gaining a
basic understanding of the role of the unstrained state, or
reference configuration, in the definition of the strain tensors.
Having a firm conceptual grasp of the role of the unstrained state
in the definition of the strain tensors is imperative for
understanding the behaviour of pre-stressed materials under finite
deformations in high-stress applications, such as, for example, in
the electromagnetic rail gun~\cite{McCorkleEMGunPaper}.

\section{Introduction}

The theory of stresses in rotating cylinders and disks is of great
importance in practical applications such as rotating machinery,
turbines and generators,  and wherever large rotational speeds are
used. In a previous work~\cite{RotatingStrainReport}, I gave a
detailed treatment of stresses in a rotating elastic cylinder.
This is a classic problem that is treated in many texts on linear
elasticity
theory~\cite{LLelasticity1970,Nadai1950,Sechler1952,Timoshenko1970,VolterraGaines1971}.
These treatments linearize the strain tensor in gradient of the
displacement field, assuming that these (dimensionless) gradients
are small. I point out in Ref~\cite{RotatingStrainReport} that for
large angles of rotation the quadratic terms (in displacement
gradient in the definition of strain) are as large as the linear
terms, and consequently, these quadratic terms cannot be dropped.
In Ref~\cite{RotatingStrainReport}, I provide an alternative
derivation of stresses in an elastic cylinder that relies on
transforming the problem from an inertial frame (where Newton's
second law is valid) to the co-rotating frame of the
cylinder--where the displacements gradients are small. During the
course of that solution, I had to transform the Lagrangian and
Eulerian strain tensors of finite elasticity to the (non-inertial)
co-rotating frame of reference of the cylinder, which is a moving,
accelerated frame. This work required the detailed understanding
of the transformation properties of the Lagrangian and Eulerian
strain tensors.

The standard derivation of these strain tensors is done with the
help of shifter tensors~\cite{Eringen1962,Narasimhan1992}. Shifter
tensors are often defined in terms of inner products of basis
vectors that are located at two different spatial
points~\cite{Eringen1962,Narasimhan1992}.  For me, inner products
between vectors at two different points is an unpalatable
operation, even in Euclidean space. In order to compute the inner
product between two vectors, the vectors must first be parallel
transported to the same spatial point (unless we are using
Cartesian coordinates, in which case the derivation becomes
coordinate specific).

In other treatments, where shifter tensors are not employed in the
derivation of strain tensors, convected (moving) coordinates are
used, see for
example~\cite{murnaghan19937,GreenZerna1954,Sokolnokoff1964,Flugge1972}.
When using convected (moving) coordinates,  the coordinates of the
initial undeformed point and the deformed point are the same, but
the basis vectors change during deformation. In derivations of
strain tensors using convected coordinates, vectors and tensors
are associated with a given point in the convected (moving)
coordinate system, rather than being associated with a point in
the underlying (inertial) space.  Tensors are absolute geometric
objects, and they should properly be associated with a point in
the underlying space, and not a given coordinate, e.g., in moving
coordinates.

In this work, I avoid the unpalatable features of the strain
tensor derivation mentioned in the above two paragraphs. I derive
the strain tensors using the concept of absolute tensors, where a
tensor is associated with a point in the space--rather than the
coordinates in a given (moving) coordinate system.  I provide a
coordinate independent derivation of the Lagrangian and Eulerian
strain tensors, where I keep track of the positions of the basis
vectors. The derivation necessarily uses two-point (and
three-point)
tensors~\cite{Eringen1962,Narasimhan1992,Ericksen1960,Synge1960}.
The derivation is based on standard concepts in differential
geometry: the parallel propagator (a two-point tensor), the world
function (a two-point scalar), and the displacement vector field
(a three-point tensor).  This derivation makes clear the
transformation properties of the strain tensors under coordinate
transformations from one frame of reference to a second frame that
is moving and accelerated (with respect to the first frame).

The derivation below of the Eulerian and Lagrangian strain tensors
makes the transformation properties (e.g., to a moving frame)
clear.  Furthermore, this derivation makes the role of the
reference (unstrained) configuration more clear in the definition
of the strain tensors. Clarifying this role is of importance for
applying finite deformation theory to pre-stresssed materials,
which are capable of withstanding higher-stress applications, such
as in rotating
machinery~\cite{RotatingStrainReport,McCorkleEMGunPaper}. Finally,
the derivation presented here allows the generalization of the
definition of strain tensors to the realm where general relativity
applies~\cite{Maugin1973,GambiEtAl1989}.

\section{Geometric Background}

In Euclidean space, a vector can be trivially parallel propagated
in the sense that after a round trip the vector still points in
the same direction.   In Riemannian space, the parallel displaced
vector is not equal to itself after the round trip parallel
displacement.   In this sense, in Euclidean space we need not
distinguish the position of a vector because ``it always points in
the same direction under parallel displacement"--even though its
components may be different from point to point because the basis
vectors, onto which we project the vector, point in a different
direction from point to point. So, in Euclidean space the parallel
displaced (physical) vector (a geometric object) is thought to
point in the same physical direction. In Riemannian space,
however, the situation is quite different.  In Riemannian space,
when a vector is parallel displaced it will in general point in a
different direction.  The physical test is to parallel displace
the vector along a curve that returns to the starting point. If
there is non-zero curvature, as measured by the Riemann curvature
tensor, then upon returning to its starting point the vector
components will be different than the initial vector components at
the starting point.  So, in Reimannian space, it is imperative to
specify the position of a vector.  In Euclidean space, appropriate
to material deformations, I also keep track of the position of a
vector. This additional care in Euclidean space, together with the
transformation properties of the world function, leads to a
clearer understanding of the transformation properties of the
Lagrangean and Eulerian strain tensors, under transformations from
one system of coordinates to another that is in relative motion (a
moving frame).

In this section I briefly review the fundamental geometric
quantities that naturally arise in discussion of deformation, but
which are not usually discussed in this context. These quantities
are the world function (or fundamental two-point scalar of the the
space), the parallel propagator, and the position vector. This
section will also serve to define my notation.  Each of the
quantities mentioned are examples of a class of geometric object
objects known as two-point tensors, which occur naturally in the
discussion of deformations. I have found useful discussions of
general tensor calculus in Synge and Schild~\cite{SyngeSchild} and
Synge~\cite{Synge1960}, and discussions oriented toward
deformation theory in the Appendix by Ericksen in Treusdell and
Toupin~\cite{Ericksen1960}, and in
Narasimhan~\cite{Narasimhan1992}, Eringen~\cite{Eringen1962}, and
Eringen~\cite{Eringen1971}.  In particular, discussion of
two-point tensors can be found in Synge~\cite{Synge1960},
Ericksen~\cite{Ericksen1960}, Narasimhan~\cite{Narasimhan1992} and
Eringen~\cite{Eringen1962}.

\subsection{The World Function}

The world function was initially introduced into tensor calculus
by Ruse~\cite{Ruse1931a,Ruse1931b}, Synge~\cite{Synge1931}, Yano
and Muto~\cite{YanoandMuto1936}, and Schouten~\cite{Schouten1954}.
It was further developed and extensively used by Synge in
applications to problems dealing with measurement theory in
general relativity~\cite{Synge1960}. An application of the world
function to problems of navigation and time transfer can be found
in Ref.~\cite{Bahder2000A}. Compared to the enormous attention
given to tensors, the world function has been used very little by
physicists. Yet, when geometry plays a central role, such as in
deformation theory, the world function is helpful to clarify and
unify the underlying geometric concepts.  The world function is
simply one-half the square of the distance between two points in
the space.  In applications to relativity and 4-dimensional
space-time, the space-time is often taken as a general (curved)
pseudo-Riemannian space~\cite{Synge1960}.  In applications to
deformation of materials, we are concerned with a Euclidean
three-dimensional space.  However, for understanding the
transformation properties of displacement vectors and strain
tensors, it is helpful to use the concept of world function in a
Euclidean 3-dimensional space described by curvilinear coordinates
$x^i$, $i=1,2,3$, with a metric $g_{ij}$, which in general is a
function of position.

Consider two points in a general Riemannian space, $P_1$ and
$P_2$, connected by a unique geodesic path (a straight line in
Euclidean space) $\Gamma$, given by $x^i(u)$, $i=1,2,3$, where
$u_1 \le u \le u_2$, and $x^i(u)$ are curvilinear coordinates of
the path. The coordinates of point $P_1= \{ x^i_1 \}$ and point
$P_2= \{x^i_2 \}$. In general, a geodesic is defined by a class of
special parameters $u^\prime$, $u \cdots$, that are related to one
another by linear transformations $u^\prime = a u + b$, where $a$
and $b$ are constants.  Here, $u$ is a particular parameter from
the class of special parameters that define the geodesic $\Gamma$,
and $x^i(u)$ satisfy the geodesic equations
\begin{equation}
\frac{d^2 x^i}{du^2}+ \Gamma^i_{jk} \frac{dx^j}{du} \frac{dx^k}{du}
=0
\label{GeodesicDiffEq}
\end{equation}
Using Cartesian coordinates   $z^k$ (rather than general
curvilinear coordinates $x^k$) in Euclidean space, the Christoffel
symbol $\Gamma^i_{jk}=0$, and the solution of
Eq.~(\ref{GeodesicDiffEq}) is simply
\begin{equation}
z^i(u) = z_1^\alpha + \frac{u-u_1}{u_2 -u_1} \, (z^\alpha_2 - z^\alpha_1)
\end{equation}
where $u_1 \le u \le u_2$,  $i=1,2,3$ and the Cartesian
coordinates of points $P_1$ and $P_2$ are $z^\alpha_1$ and
$z^\alpha_2$, respectively.  In a general Riemannin space, the
world function between point $P_1$ and $P_2$ is defined as the
integral along $\Gamma$ in arbitrary curvilinear coordinates $x^i$
by
\begin{equation}
\Omega(P_1,P_2) = \frac{1}{2} (u_2 - u_1) \int^{u_2}_{u_1} \, g_{ij} \frac{dx^i}{du}
\frac{dx^j}{du} \, du
\label{WorldFunctionDef}
\end{equation}
The value of the world function has a simple geometric meaning: it is
one-half the distance between points $P_1$ and $P_2$. Its value
depends only on the eight coordinates of the points $P_1$ and
$P_2$. The value of the world function in Eq.\
(\ref{WorldFunctionDef}) is independent of the particular special
parameter $u$ in the sense that under a transformation from one
special parameter $u$ to another, $u^\prime$, given by $u=a
u^\prime + b$, with $x^i(u)=x^i(u(u^\prime))$, the world function
definition in Eq.\ (\ref{WorldFunctionDef}) has the same form
(with $u$ replaced by $u^\prime$).

The world function is {\it the} fundamental two-point invariant
that characterizes the space. It is invariant under independent
transformation of coordinates at $P_1$ and at $P_2$.  For a given
space, the world function between points $P_1$ and $P_2$ has the
same value independent of the coordinates used, which makes it a
useful coordinate independent quantity. In Euclidean space, using
Cartesian coordinates, the world function has the simple form
\begin{equation}
\Omega(z^i_1,z^j_2) = \frac{1}{2}\, \delta_{ij} \, \Delta z^i  \,  \Delta z^j
\label{CartesianEuclideanWorldFunction}
\end{equation}
where $\delta_{ij}$ is the Euclidean metric with only non-zero
diagonal components $(+1,+1,+1)$, and $ \Delta z^i  = (z_2^i -
z_1^i)$, $i=1,2,3$, where $z_1^i$ and $z_2^i$ are the Cartesian
coordinates of points $P_1$ and $P_2$, respectively. (I use the
convention that all repeated indices are summed, unless otherwise
stated.)

The world function has a number of interesting properties,
see Synge~\cite{Synge1960}. Calculations of the world function for
spaces other than Euclidean spaces, namely four-dimensional
space-time, can be found in
Refs.~\cite{Synge1960,Bahder2000A,John1984,John1989,Buchdahl79}.
In what follows, I restrict myself to a three-dimensional space.
By transforming to a new system of coordinates, say spherical
coordinates,
\begin{eqnarray}
x & = & r \, \cos \theta \,\, \cos \phi  \label{eq1} \\
y & = & r \, \cos \theta \,\, \sin \phi \label{eq2} \\
z & = & r \, \cos \theta  \label{eq3}
\end{eqnarray}
the world function in Eq.~(\ref{CartesianEuclideanWorldFunction})
can be expressed as a function of spherical coordinates of point
$P_1=(r_1,\theta_1,\phi_1)$, and $P_2=(r_2,\theta_2,\phi_2)$.

\begin{figure}[htbp]
\centerline{\epsfsize=6.0cm \epsfbox{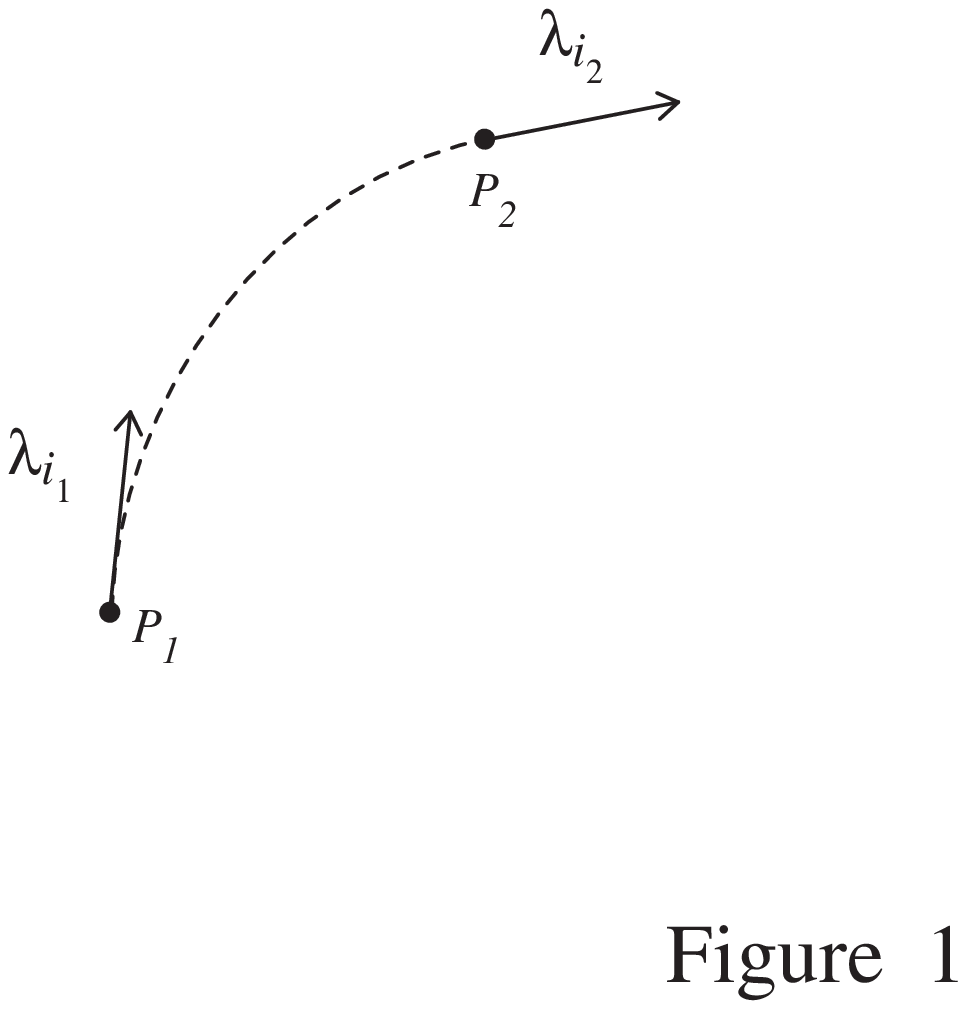}}
\caption{A geodesic path is shown in 3-dimensions, with tangent
unit vector at the ends.} \protect
\label{path}
\end{figure}

Consider a geodesic given by Eq.~(\ref{GeodesicDiffEq}) in a
general 3-dimensional Riemannian space. The covariant derivatives
of the world function have two important properties:
\begin{eqnarray}\label{geodesic}
\frac{\partial \, \Omega(P_1,P_2)}{\partial \, x^i_2} & = &
(u_2-u_1)\left(  g_{ij} \,  \frac{dx^j}{du} \right)_{P_2}   =  L \, \lambda_{i_2}  \label{endPoint}\\
\frac{\partial \, \Omega(P_1,P_2)}{\partial \, x^i_1} &  = &
-(u_2-u_1)\left(  g_{ij} \,  \frac{dx^j}{du} \right)_{P_1}   =  -L
\, \lambda_{i_1}  \label{beginPoint}
\end{eqnarray}
where
\begin{equation}\label{length}
L=\left[ 2 \, \Omega(P_1,P_2) \right]^{1/2} = \int^{P_2}_{P_1}\,ds
= \,  \int^{u_2}_{u_1}\, \left[ g_{ij} \, \frac{dx^i}{du}\, \frac{dx^j}{du} \,
\right]^{1/2} \, du
\end{equation}
is the length of the geodesic between $P_1$ and $P_2$, $g_{ij}$ is
the metric in coordinates $x^i$, and  $\lambda_{i_1}$ and
$\lambda_{i_2}$ are components of the unit tangent vectors at end
points $P_1$ and $P_2$ (assuming non-null
geodesics~\cite{Synge1960}):
\begin{eqnarray}
\lambda_{i_1} & = & \left( g_{ij} \, \frac{dx^j}{ds} \right)_{P_1} \label{lam1} \\
\lambda_{i_2} & = & \left( g_{ij} \, \frac{dx^j}{ds} \right)_{P_2}
\label{lam2}
\end{eqnarray}
where the relation between parameter $u$ and arc length $s$ is
given by Eq.~(\ref{length}). In Eq.~(\ref{endPoint}) and
(\ref{beginPoint}), the covariant partial derivatives with respect
to $x^i_1$ and $x^i_2$ are done with respect to the coordinates of
points $P_1$ and  $P_2$. See Fig.~(\ref{path}).

For the special case of interest in deformation of materials, the
space is three-dimensional Euclidean, the geodesic is a straight
line, and the vectors $\lambda_{i_1}$ and $ \lambda_{i_2}$ are
colinear, although I still consider them as existing at distinct
points. Using Cartesian coordinates and the explicit form of the
world function in Eq.~(\ref{CartesianEuclideanWorldFunction}),
Eq.~(\ref{endPoint}) and (\ref{beginPoint}) take the form
\begin{eqnarray}\label{CartesianDerivative}
\frac{\partial \, \Omega(P_1,P_2)}{\partial \, z^i_2} & = & \equiv
\Omega_{i_2} = \delta_{ij} \, \left( z^j_2-z^j_1\right) =
 L \,\lambda_{i_2}  \label{ZendPoint}\\
\frac{\partial \, \Omega(P_1,P_2)}{\partial \, z^i_1} & = & \equiv
\Omega_{i_1} = -\delta_{ij} \, \left( z^j_2-z^j_1\right) =
 -L \, \lambda_{i_1}  \label{ZbeginPoint}
\end{eqnarray}
where I used Synge's short-hand notation for the components of the
covariant partial derivatives by putting subscripts on the indices
to indicate which coordinates were differentiated.  This
short-hand notation is particularly convenient to show the
transformation properties of the world function and to indicate
the spatial location of vectors and tensors. For example, the
quantity $\Omega_{i_2}$ transforms as a vector under coordinate
transformations at $P_2$ and as a scalar under coordinate
transformations at point $P_1$. The quantity $\lambda_{i_2}$ is a
vector located at point $P_2$. Note that by virtue of their
definitions in the left side of Eq.~(\ref{ZendPoint}) and
(\ref{ZbeginPoint}), the right sides are two-point tensors, whose
components are functions of coordinates at point $P_1$ and $P_2$.
For example, the right side of Eq.~(\ref{ZbeginPoint}) is a
product of a two-point scalar $L$, and a one point vector,
$\lambda_{i_1}$ at $P_1$.

\subsection{Parallel Propagator}

Given a vector with components $v^{i_1}$ at point $P_1$, the
vector is said to be parallel propagated from $P_1$ to $P_2$ along
a geodesic curve $C$ specified by $x^i(u)$, $u_1 \le u \le u_2$,
where $P_1=  \{ x^i(u_1) \}$ and $P_2=\{ x^i(u_2) \}$,  when its
covariant derivative is zero along this curve:
\begin{equation}\label{ParallelPropagation}
\frac{dv^i}{d u} + \Gamma^i_{jk}\, v^j \, \frac{dx^k}{d u}
= 0
\end{equation}
Equation~(\ref{ParallelPropagation}) is a mapping: given the
components of a vector, $v^{i_1}$ at point $P_1$, we obtain the
components $v^{i_2}$ of the parallel transported vector at point
$P_2$ by solving Eq.~(\ref{ParallelPropagation}).  It is
convenient to define a two-point tensor, $g^{i_2}_{~ j_1}$, called
the parallel propagator~\cite{Synge1960}, which gives the
components of a vector under parallel translation of the vector
from point $P_1$ to point $P_2$.  Given a vector with components
$v^{i_1} $ at point $P_1$, the propagator  $g^{i_2}_{~ j_1}$
relates the components of this vector at $P_1$ to the components
$v^{i_2}$ of this same vector after parallel translation to point
$P_2$
\begin{equation}\label{propagatorDef}
 v^{i_2} = g^{i_2}_{~~ i_1} \, v^{i_1}
\end{equation}

In a general Riemannian space, the components of the vector at
point $P_2$ depend on the path of parallel translation from $P_1$
to $P_2$, in the sense that the path must be a geodesic by the
definition of the parallel propagator. However, in Euclidean space
these components are completely path independent; the components
depend only on the end points $P_1$ and $P_2$.

A vector is considered as a geometric object, which means that it
is independent of coordinate system.  In a Riemannian space, under
the operation of parallel propagation a vector changes in such a
way that its magnitude stays the same but its absolute direction
can change because of the curvature of the
space~\cite{Krauss1975}. The direction of the parallel propagated
vector is of course referred to the local basis vectors. That the
vector direction changes under parallel translation can be
understood by taking a vector at point $P$ and parallel
translating it over a curve that returns to point $P$. When
compared at point $P$, the components of the initial vector and
the round-trip-parallel-transported vector will (in general) be
different.  It is in this sense that a vector changes its
direction under parallel transport.

As mentioned above, the change in the vector that results under
parallel transport depends on the path of parallel propagation (a
geodesic). Two vectors that are parallel propagated along the same
path will maintain the angle between them along the path.

In a Euclidean space, a vector (the geometric object) is
considered to be  unchanged when parallel propagated.  The only
thing that happens is that the components of the vector on the
local basis must change according to what is required to keep the
vector ``pointing in the in same direction".

In Euclidean space, the parallel propagator in Cartesian
coordinates is trivial--its components are just the components of
a delta function.  The components of a vector at point $P_1$ are
related to the components of the same vector parallel translated
to point $P_2$ by the propagator (whose components are given in a
Cartesian coordinate basis):
\begin{equation}\label{CartesianPrarllelPropDef}
\delta^{i_2}_{~~ j_1} = \left\{
\begin{array}{c}
  +1 \, \, \, \, \, \, i=j \\
   0 \, \, \, \, \, \, i \ne j
\end{array}
\right.
\end{equation}
Equation~(\ref{CartesianPrarllelPropDef}) agrees with our notion
from elementary geometry that in Cartesian coordinates the vector
components are constant under parallel propagation.  However,
using the parallel propagator in Cartesian coordinates, I can, for
example, compute the propagator $g^{i_2}_{~~ j_1}$ in curvilinear
coordinates $x^i=(r,\theta,\phi)$ given in
Eq.~(\ref{eq1})--(\ref{eq3}), by the two-point tensor
transformation rule
\begin{equation}\label{propagatorTransformation}
 g^{i_2}_{~~ j_1} = \frac{\partial \,
x^i(P_2)}{\partial \, z^m} \, \frac{\partial \, z^n(P_1)}{\partial
\, x^j} \, \delta^{m_2}_{~~ n_1}
\end{equation}
The parallel propagator $ g^{i_2}_{~~ j_1}$ is a two-point tensor
because it transforms as a vector under  coordinate transformation
at point $P_1$ and under coordinate transformation at \hbox{point
$P_2$}.

In Cartesian coordinates, when the points are made to coincide,
$P_2\rightarrow P_1$, the propagator reduces to a Kronecker delta
at point $P_1$: $\delta^{m_2}_{~~ n_1} \rightarrow \delta^m_{~~
n}(P_1)$.  In general curvilinear coordinates, when the points
$P_1$ and $P_2$ coincide,  the parallel propagator reduces to the
mixed components of the mertic tensor $g^{i_2}_{~~ j_1}$:
\begin{equation}
\lim_{P_2 \rightarrow P_1} g^{i_2}_{~~ j_1} \rightarrow g^i_{~
j}(P_1) \, \, \, \, \, \mbox{(metric at $P_1$)}
\end{equation}
The mixed components of the metric tensor at $P_1$, $g^i_{~
j}(P_1) \equiv g^{ik}(P_1) \, g_{kj}(P_1)$, are a Kronecker
\hbox{delta--a unit} tensor whose components are the same in all
systems of coordinates. Indices can be lowered on two-point
tensors using the appropriate metric. For example, the index $i$
of the propagator $g^{i_2}_{~~ j_1}$ can be lowered by using the
metric tensor at point $P_2$:
\begin{equation}\label{PropagatorLowerIndicies}
g_{k_2 j_1} = g_{ki}(P_2) \,  g^{i_2}_{~~ j_1}
\end{equation}
When the points are made to coincide, $P_2\rightarrow P_1$, the
covariant components of the propagator becomes the covariant
components of the metric tensor at $P_1$, $g_{k_2 j_1} \rightarrow
g_{kj}(P_1)$, where $g_{kj}(P_1)$ is the metric at $P_1$.

The covariant derivatives of the world function $\Omega(P_1,P_2)$
between points $P_1$ and $P_2$ are related to the parallel
propagator by~\cite{Synge1960}
\begin{eqnarray}
\Omega_{i_1 j_1} & = & g_{i_1 j_1} \,\,\,\, \mbox{(metric at $P_1$)} \label{OmProp1} \\
\Omega_{i_1 j_2} & = & \Omega_{j_2 i_1} = -g_{i_1 j_2} = -g_{j_2 i_1} \,\,\,\, \mbox{(parallel propagator)} \label{OmProp2} \\
\Omega_{i_2 j_2} & = & g_{i_2 j_2} \,\,\,\, \mbox{(metric at
$P_2$)} \label{OmProp3}
\end{eqnarray}
Other useful properties of the parallel propagator are discussed
by Synge~\cite{Synge1960}.

\subsection{Position Vector}

The position vector occupies a central role in deformation theory.
For this reason, I discuss it in detail below.  In elementary
geometry, a point $P$ can be identified by its position vector
${\bf r}$, which can be specified in Euclidean-space Cartesian
coordinates as
\begin{equation}\label{positionVector}
{\bf r}  = z^n \, {\bf i}_n
\end{equation}
where $z^n$ are the Cartesian components of the vector ${\bf r}$
and also the Cartesian coordinates of the point $P$.  In terms of
general  coordinate basis vectors ${\bf e}_n = \partial / \partial
x^n$ associated with the curvilinear coordinates $x^n$, the vector
${\bf r}$ is given by
\begin{equation}\label{positionVector2}
{\bf r}  = z^n \, A^m_{~~n}(P) \, {\bf e}_m(P) = \zeta^ m \, {\bf
e}_m(P)
\end{equation}
The position vector is a geometric object at point $P$.  Among all
basis vectors, the Cartesian basis vectors ${\bf i}_n$ are unique
in that they are usually not associated with a particular spatial
point. However, when we express these Cartesian basis vectors in
terms of curvilinear basis vectors ${\bf e}_n$, then we must
imagine that these basis vectors exist at a particular point $P$.
Hence, the transformation between the Cartesian basis vector ${\bf
i}_m$  and curvilinear coordinate basis vectors ${\bf e}_m$ at
point $P$ associated with coordinates $x^i$ is given by
\begin{equation}\label{CartesianBasisTransform}
{\bf i}_n(P) = A^m_{~~n}(P) \, {\bf e}_m(P)
\end{equation}
where the matrix $A^m_{n}(P)$ depends on the coordinates of point
$P$:
\begin{equation}\label{Amndef}
A^m_{~~n}(P) = \frac{\partial \, x^m}{\partial \, z^n}(P)
\end{equation}

In Cartesian coordinates, the components of the vector ${\bf r}$
are simply the Cartesian coordinates $z^n$ of point $P$. The three
numbers $(z^1,z^2,z^3)$ transform as the components of a vector
under orthogonal coordinate transformations.   Note that in
curvilinear coordinates, the components of the position vector,
$\zeta^m$, are not the curvilinear coordinates of point $P$. Also,
note that the position vector ${\bf r}$ of point $P$ has a
magnitude equal to the Euclidean length from the origin of
coordinates, say point $O$, to point $P$.  The position vector of
point $P$ is a geometric object at point $P$, however, it also
depends on the point $O$. This dependence on point $O$ is
coordinate independent. Therefore, the position vector of point
$P$ is a two-point tensor; it depends on point $P$ and on point
$O$. The transformation properties of the position vector are that
of a scalar when a change of coordinates is made at point $O$ and
the transformation is that of a vector when coordinates at point
$P$ are changed.

In a Riemannian (a generalization of Euclidean space) space, the
components of the position vector $r_i(P)$ at point $P$ can be
defined in terms of the covariant derivative of the world function
\begin{equation}\label{WorldFunctionCovariantDeriv}
r_{i_P} = \frac{\partial \, \Omega(O,P)}{\partial \, z^i_P} \equiv
\Omega_{i_P}(O,P) = \left[ 2 \, \Omega_{i_P}(O,P) \right]^{1/2} \,
\hat{r}^i(P)
\end{equation}
where $ \hat{r}^{i_P}$ is a unit vector at point $P$ and $\left[ 2
\, \Omega_{i_P}(O,P) \right]^{1/2} $ is the length of the geodesic
from point $O$ to point $P$. For the case of Euclidean space,
$\left[ 2 \, \Omega_{i_P}(O,P) \right]^{1/2}$  is the length of
the straight line $\overline{OP}$.
Equation~(\ref{WorldFunctionCovariantDeriv}) shows explicitly that
the position vector, $r_{i_P}$, is a two-point tensor.

\subsection{Displacement Vector}

Consider an elastic body that undergoes a finite deformation in
time. The deformation can be specified by a flow function or
displacement mapping function
\begin{equation}\label{defMapFcn}
z^k = z^k(Z^m,t)
\end{equation}
where the coordinates $z^k$ (here taken to be Cartesian) of a
particle at point $Q$ at time $t$ are given in terms of the
particle's coordinates $Z^k$ of point $P$ in some reference state
(configuration) at time $t=t_o$, so that $z^k(Z^m,t_o)=Z^k$.  I
assume the deformation mapping function has an inverse, which I
quote here for later reference
\begin{equation}\label{defMapFcnInverse}
Z^m = Z^m(z^k,t)
\end{equation}
I assume that both the coordinates $z^k$ and  $Z^m$ refer to the
same Cartesian coordinate system~\cite{FootNote1}.

In deformation theory, the initial position of the particle in the
medium  at point $P$ is specified by a vector
\begin{equation}\label{InitialPositionVector}
{\bf R}(P)= Z^mk \, {\bf i}_k(P)
\end{equation}
and the final position is specified by a position vector
\begin{equation}\label{FinalPositionVector}
{\bf r}(Q)= z^k \, {\bf i}_k(Q)
\end{equation}
where the quantities ${\bf i}_k(P)$ and ${\bf i}_k(Q)$ are the
Cartesian basis vectors at point $P$ and point $Q$, respectively.
Conventionally, the deformation of a medium is described by
specifying the displacement ``vector field", which is defined as a
difference of these two position vectors.  However, the basis
vectors ${\bf i}_k(P)$ and ${\bf i}_k(Q)$ are at different points
in the space.   Since vectors can be subtracted only if they are
at the same point,   I must parallel translate  ${\bf i}_k(P)$ to
point $Q$, or, parallel translate ${\bf i}_k(Q)$ to point $P$.
Depending on which mapping I choose, I arrive at the Eulerian or
the Lagrangian displacement vector.

\begin{figure}[htbp]
\centerline{\epsfsize=6.0cm \epsfbox{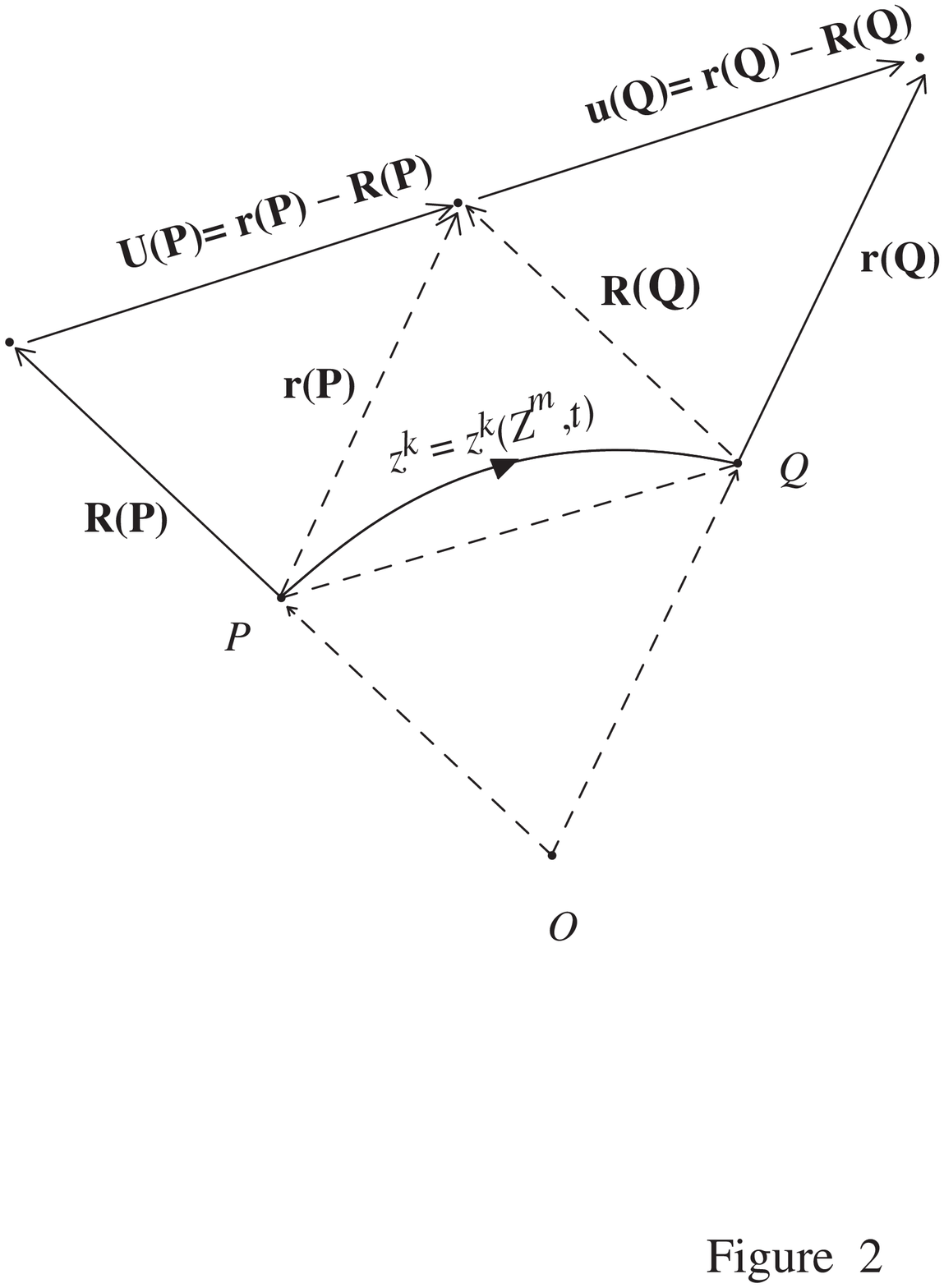}}
\caption{The initial and final position vectors, ${\bf R}(P)$ and
${\bf r}(Q)$, respectively, are shown as well as their respective
parallel translated vectors, ${\bf R}(P)$ and ${\bf r}(Q)$. Also,
shown by a solid curve is the actual displacement path of a
representative particle of the medium, labeled by
$z^k=z^k(Z^m,t)$. The dashed straight line is the line (geodesic)
connecting the initial and final particle positions. The Eulerian
displacement vector is ${\bf u} = {\bf r}(Q) - {\bf R}(Q)  $,
which is the difference of two position vectors at point $Q$. The
Lagrangean displacement vector, ${\bf U} = {\bf r}(P) - {\bf R}
(P)$, is the difference of two position vectors at point $P$. }
\protect \label{displacementVectors}
\end{figure}

First, I parallel translate vector ${\bf R}(P)$ to point $Q$, and
then subtract the vectors  at point $Q$, see
Fig.~(\ref{displacementVectors}). This procedure defines the
components of the Eulerian displacement vector at point $Q$:
\begin{equation}\label{EulerianDisplacementVectorDef}
{\bf u}(Q)= {\bf r}(Q) - {\bf R}(Q)
\end{equation}
This  Eulerian displacement vector in
Eq.~(\ref{EulerianDisplacementVectorDef}) is often called the
displacement vector in the spatial
representation~\cite{Eringen1962,Narasimhan1992,Spencer1980}.
Alternatively, I can parallel translate the vector ${\bf r}(Q)$ to
point $P$, and then subtract the vectors at point $P$.  This
procedure defines the Lagrangian displacement vector at point $P$:
\begin{equation}\label{LagrangianDisplacementVectorDef}
{\bf U}(P)= {\bf r}(P) - {\bf R}(P)
\end{equation}
This  Lagrangian displacement vector in
Eq.~(\ref{LagrangianDisplacementVectorDef}) is often called the
displacement vector in the material
representation~\cite{Eringen1962,Narasimhan1992,Spencer1980}.
Equations~(\ref{EulerianDisplacementVectorDef}) and
(\ref{LagrangianDisplacementVectorDef}) show that these two
vectors are actually referred to basis vectors at different
points. In fact, the two vectors ${\bf u}(Q)$ and ${\bf U}(P)$ are
related by parallel translation. In a Euclidean space, these
vectors  are the same geometric objects but they are expressed in
terms of basis vectors located
at different positions 

In order to further clarify the transformation properties of these
two displacement vectors, I use the  position vector  as discussed
in the previous section.  Consider the deformation mapping
function in curvilinear coordinates, $x^k(X^m,t)$.  This function
specifies the coordinates $x^k$  (point $Q$) of a particle at
current time $t$ in terms of the coordinates $X^k$ (point $P$) of
the particle in the reference configuration at time $t=t_o$, so
that
\begin{equation}\label{DefMapFn}
x^k(X^m,t_o)=X^k
\end{equation}
In addition, there exists a straight line (a geodesic) $\Gamma$
connecting the points $P$ and $Q$.

The covariant components of the position vector of point $P$,
${\bf R}(P)=R^n \, {\bf e}_n(P)$,  in curvilinear coordinates $x^i$ are
given by (see Eq.~(\ref{WorldFunctionCovariantDeriv}))
\begin{equation}\label{Rcomponents-P}
R_{i_P} = \frac{\partial \, \Omega(O,P)}{\partial \, x^i_P} \equiv
\Omega_{i_P}(O,P) =  \left[ 2 \,  \, \Omega(O,P) \right]^{1/2} \,
\hat{R}_{i_P}
\end{equation}
where $ \hat{R}_{i_P}$ are the components of the unit vector at
point $P$ tangent to $\Gamma$  that connects point $P$ and $Q$.
Similarly, the covariant components of vector ${\bf r}(Q)=r^n \,
{\bf e}_n(Q)$ in curvilinear coordinates $x^i$ are given by
\begin{equation}\label{r-components-Q}
r_{i_Q} = \frac{\partial \, \Omega(O,Q)}{\partial \, x^i_Q} \equiv
\Omega_{i_Q}(O,Q)  =  \left[ 2 \, \Omega(O,Q) \right]^{1/2} \,
\hat{r}_{i_Q}
\end{equation}
where $\hat{r}_{i_Q}$ are the components of the unit vector at
point $Q$ tangent to $\Gamma$. From Eq.~(\ref{Rcomponents-P}) and
(\ref{r-components-Q}) it is clear that both $R_{i_P}$ and
$r_{i_Q}$ are two-point tensor objects. The quantity $R_{i_P}$
depends on point $O$ and $P$ and transforms as a vector under
coordinate transformations at $P$ and as a scalar under coordinate
transformations at point $O$.  The quantity $r_{i_Q}$ transforms
as a vector under coordinate transformations at point $Q$ and as a
scalar under coordinate transformations at point $O$.

The  components of the Eulerian displacement vector in
Eq.~(\ref{EulerianDisplacementVectorDef}) (at point $Q$) are
defined in terms of the components of $R^{i_P}$ parallel
translated to point $Q$:
\begin{eqnarray}
R^{i_Q} & = &  g^{i_Q}_{~~ j_P} \, R^{j_P}  \label{R-translated} \\
       &  = &  g^{i_Q}_{~~ j_P}
        \, \, \Omega^{j_P}(O,P)  \label{R-translated2}
\end{eqnarray}
where $\Omega^{j_P}(O,P)=g^{jk}(P) \, \Omega_{k_P}(O,P)$, and
$g^{jk}(P)$ is the metric at point $P$ in coordinates $x^i$.
The contravariant components of the Eulerian displacement vector
in Eq.~(\ref{EulerianDisplacementVectorDef}) are given by
\begin{equation}\label{EulerianDispComponents}
u^{i_Q} = \Omega^{i_Q}(O,Q) -   g^{i_Q}_{~~ j_P} \,
\Omega^{j_P}(O,P)
\end{equation}
Similarly, the components of the Lagrangian displacement vector
are given by
\begin{equation}\label{LagrangianDispComponents}
 U^{i_P} = g^{i_P}_{~~ j_Q} \, \Omega^{j_Q}(O,Q)  -
\Omega^{i_P}(O,P)
\end{equation}
The vectors whose components are $U^{i_P}$ and $u^{i_Q}$, are
related by parallel transport along the geodesic $\Gamma$
connecting $P$ and $Q$ (not along the particle displacement line
given by Eq.(\ref{defMapFcn})). Transporting $U^{i_P}$ to point
$Q$
\begin{eqnarray}
U^{i_Q} & = &   g^{i_Q}_{~~ k_P}  \, U^{k_P} \label{Utransported1} \\
       & = &   g^{i_Q}_{~~ j_P} \, \left[ g^{j_P}_{~~ k_Q} \, \Omega^{k_Q}(O,Q) - \Omega^{j_P}(O,P)
       \right] \label{Utransported2} \\
       &  = &  \delta^i_{~k}(Q) \Omega^{k_Q}(O,Q) - g^{i_Q}_{~~ j_P}
       \,  \Omega^{j_P}(O,P)  \label{Utransported3} \\
        & = &   u^{i_Q}
\end{eqnarray}
where in the transition from Eq.~(\ref{Utransported2}) to
(\ref{Utransported3}) I have used the identity satisfied by the
parallel propagator:
\begin{equation}\label{deltaDef}
 \delta^i_{~k}(Q) = g^{i_Q}_{~~ j_P} \,  g^{j_P}_{~~ k_Q}
\end{equation}
where $ \delta^i_{~k}(Q)$ is a unit tensor (delta function) at
point $Q$. Equation~(\ref{deltaDef}) is the statement that
parallel propagation of a vector has an inverse, so the result
that $U^i(P)$ and $u^i(Q)$ are related by parallel transport is
true in both Euclidean and Reimannian spaces.

\section{Strain Tensors}

\begin{figure}[htbp]
\centerline{\epsfsize=6.0cm \epsfbox{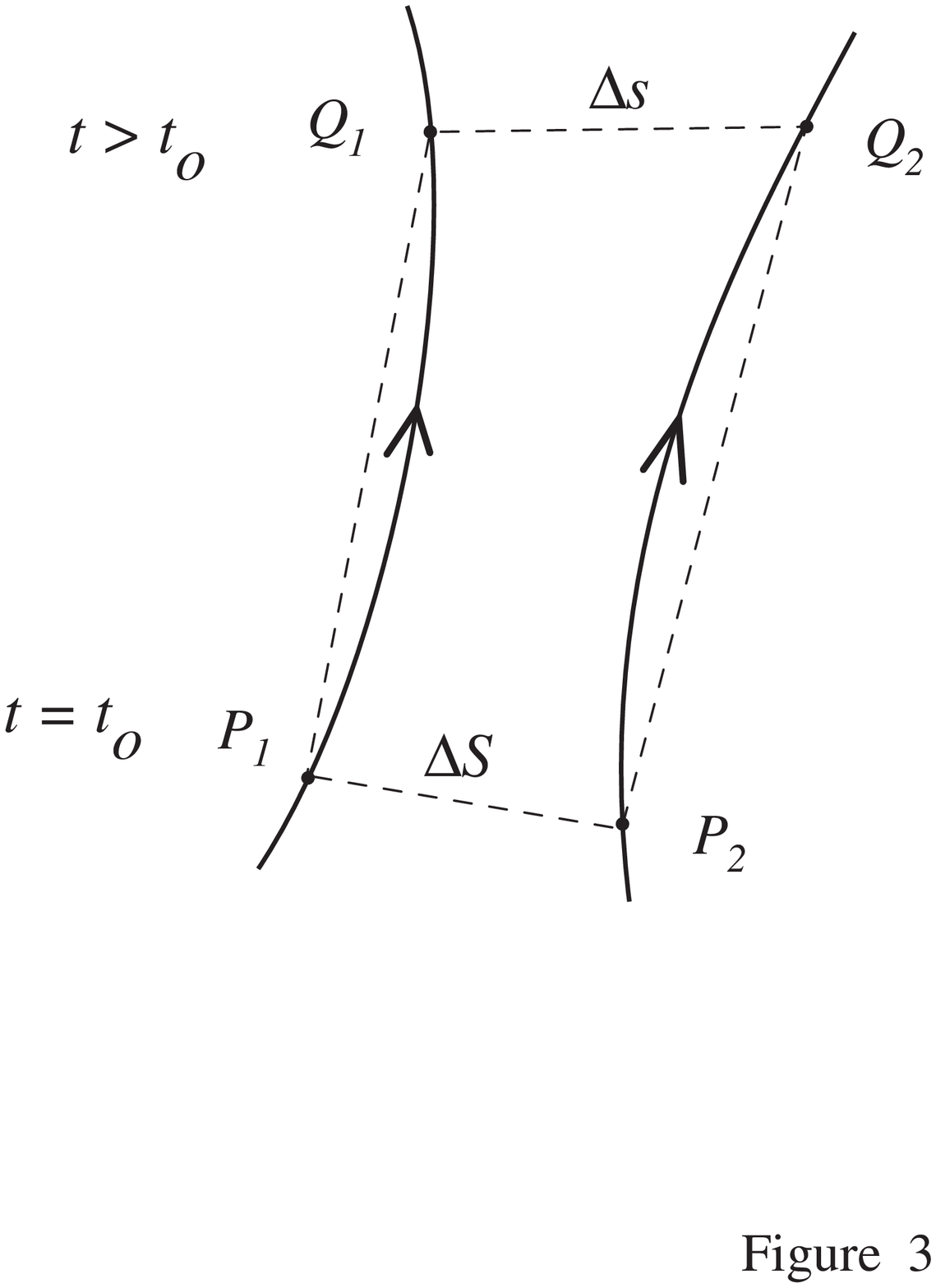}} \caption{The
position $P_1$ and $P_2$ of two particles is shown in the
reference configuration at $t=t_o$, and the positions $Q_1$ and
$Q_2$ of the same two particles is shown at later time $t>t_o$.
The path of the particles is shown in solid lines and their
displacement is shown in dashed lines.} \protect
\label{strainFig}
\end{figure}

At time $t=t_o$, consider two particles in the medium that are at
positions $P_1$ and $P_2$, respectively, and are separated by a
finite distance $\Delta S = \left[ 2 \, \Omega(P_1,P_2)
\right]^{1/2} $. At a later time $t>t_o$ these particles have
moved to new positions $Q_1$ and $Q_2$ and are separated by a
distance $\Delta s = \left[ 2 \, \Omega(Q_1,Q_2) \right]^{1/2}$,
see Fig.~(\ref{strainFig}).

As a measure of strain, I take the 4-point scalar
\begin{equation}\label{strainMeasure}
\Psi(P_1,P_2; Q_1,Q_2) \equiv (\Delta s)^2 - (\Delta S)^2 = 2 \Omega(Q_1,Q_2)  - 2 \Omega(P_1,P_2)
\end{equation}
Note that $ \Psi(P_1,P_2; Q_1,Q_2)$ depends on four points $P_1$,
$P_2$, $Q_1$, and $Q_2$, and by virtue of its definition in terms
of the world function, $\Psi$ is a true four-point scalar
invariant under separate coordinate transformations at each of
these four points.  In Cartesian coordinates, Eq.~(\ref{strainMeasure})
becomes
\begin{equation}\label{CartesianStrainTensor}
\Psi(z^i_1, z^i_2; Z^j_1, Z^j_2) = \delta_{ij} \, (z^i_1 - z^i_2) \, (z^j_1 - z^j_2))
- \delta_{ij} \, (Z^i_1 - Z^i_2) \, (Z^j_1 - Z^j_2))
\end{equation}
where $z^i_1=z^i(Z^m_1,t)$ and $z^i_2=z^i(Z^m_2,t)$ and they are
related to the reference configuration at $t=t_o$ by
\begin{equation}\label{particel1}
z^i(Z^m_1,t_o)=Z^i_1
\end{equation}
and
\begin{equation}\label{particel2}
z^i(Z^m_2,t_o)=Z^i_2
\end{equation}
and $z^i(Z^m,t)$ is the deformation mapping function in Cartesian
coordinates, given in Eq.~(\ref{defMapFcn}). If I consider the
particle positions $P_1$ and $P_2$ as separated by an
infinitesimal distance, then, by assuming continuity in the medium
and a finite time $t-t_o$, the points $Q_1$ and $Q_2$ are also
infinitesimally separated. However, because $t-t_o$ is finite, the
distance between $P_1$ and $Q_1$ is finite (not infinitesimal).
Expanding Eq.~(\ref{particel2}) about the initial position of the
first particle
\begin{equation}\label{expansion1}
z^i_2 = z^i (Z^k_1,t) + \frac{\partial \, z^i}{\partial \, Z^j}(Z^m_1,t) \,
(Z^j_2 - Z^j_1) + \cdots
\end{equation}
using $z^i_1=z^i (Z^k_1,t)$, leads to the relation between spatial
(Eulerian) coordinates $z^i$ and material (Lagrangian) coordinates $Z^k$
\begin{equation}\label{expansion2}
\Delta z^i =  \frac{\partial \, z^i}{\partial \, Z^j}(Z^m_1,t) \,
\Delta Z^j + \cdots
\end{equation}
where $\Delta z^i=  z^i_2 - z^i_1$ and $\Delta Z^i=  Z^i_2 -
Z^i_1$.  Using Eq.~(\ref{expansion2}) in
Eq.~(\ref{CartesianStrainTensor}) leads to the measure of strain
in Cartesian coordinates
\begin{eqnarray}\label{LagrangeanStrainTensor}
(\Delta s)^2 - (\Delta S)^2  & = & \left( \delta_{ij} \,
\frac{\partial \, z^i}{\partial \, Z^m} \, \frac{\partial \, z^i}{\partial \, Z^n}  -
\delta_{mn} \right) \, \Delta Z^m \,  \Delta Z^n   + \cdots \label{LagrangeanStrainTensor1} \\
  &  = &  2 \, E_{mn} \, \Delta Z^m \,  \Delta Z^n + \cdots  \label{LagrangeanStrainTensor2}
\end{eqnarray}
where the quantity in parenthesis is the Lagrangean strain tensor,
$E_{mn}$.  The higher order terms in $\Delta Z^m$ are small
because I can always choose the two initial points $P_1$ and $P_2$
arbitrarily close together. From
Eq.~(\ref{LagrangeanStrainTensor1})--(\ref{LagrangeanStrainTensor2}),
it is clear that the Lagrangean strain tensor is a two-point
tensor, depending on initial point $P$ (in the reference
configuration at $t=t_o$) and point $Q$ (at time $t$).  Note that
in Eq.~(\ref{LagrangeanStrainTensor2}) there is no restriction to
short times $t-t_o$, since I can always choose $\Delta Z^n$
sufficiently small.

The Eullerian tensor can be obtained from
Eq.~(\ref{LagrangeanStrainTensor2}) by using the fact that the
flow function in  Eq.~(\ref{defMapFcn}) has an inverse.  Using
\begin{equation}\label{gradient}
\Delta \, Z^n = \frac{\partial Z^n}{\partial z^i}(z^i,t)  \, \Delta
z^i
\end{equation}
I get the measure of strain in terms of the Eulerian strain tensor
$e_{mn}$:
\begin{eqnarray}\label{EulerianStrainTensor}
(\Delta s)^2 - (\Delta S)^2  & = & \left( \delta_{mn} -
\delta_{kl} \, \frac{\partial \, Z^k}{\partial \, z^m} \, \frac{\partial \, Z^l}{\partial \, z^n}
 \right) \, \Delta z^m \,  \Delta z^n   + \cdots \label{EulerianStrainTensor1} \\
  &  = &  2 \, e_{mn} \, \Delta z^m \,  \Delta z^n + \cdots  \label{EulerianStrainTensor2}
\end{eqnarray}

\subsection{Strain Tensor Derivation in Curvilinear Coordinates}

I return to the definition of the measure of strain given in
Eq.~(\ref{strainMeasure}).  In the reference configuration at
$t=t_o$, consider two particles at points $P_1$ and $P_2$ with
curvilinear coordinates $X_1$ and $X_2$. At a later time $t$,
these two particles are at positions $Q_1$ and $Q_2$  with
curvilinear coordinates $x_1$ and $x_2$, respectively.  Consider
the first term on the right side of Eq.~(\ref{strainMeasure}),
which is an integral along a straight line $\overline{Q_1Q_2}$:
\begin{equation}\label{FirstTerm}
\Omega(Q_1,Q_2) = \frac{1}{2} (w_2 - w_1) \int^{w_2}_{w_1} \, g_{ij}
U^i \, U^j \, dw
\end{equation}
where $U^i = dx^i(w)/dw$ and where the geodesic (straight line) is
parametrized by $x^i(w)$, with $w_1 \le w \le w_2$ and the end
points are given by $x_1=x^i(w_1)$ and $x_2=x^i(w_2)$, see
Fig.~\ref{strainFig}. The function $x^i(w)$ is a solution of the
geodesic Eq.~(\ref{GeodesicDiffEq}).  In the  case of Eulidean
space,  and assuming points $P_1$ and $P_2$ are arbitrarily close,
the geodesic in Eq.~(\ref{FirstTerm}) is a straight line given by
\begin{equation}\label{straightline}
x^i(w)=x^i_1 + \frac{w-w_1}{w_2-w_1} \, (x^i_2 - x^i_1)
\end{equation}
The flow function in  Eq.~(\ref{defMapFcn}) maps the points $P_1$
and $P_2$ into the points $Q_1$ and $Q_2$.  The points $Q_1$ and
$Q_2$ depend on time $t$. With reasonable continuity assumptions,
and the straight line given in Eq.~(\ref{straightline}) with
$U^{i_Q}= k (x^i_2-x^i_1)=k\, \Delta x^i$ and $k=(w_2-w_1)^{-1}$,
the world function in Eq.~(\ref{FirstTerm}) can be approximated by
\begin{eqnarray} \label{FirstTerm2}
\Omega(Q_1,Q_2) & = & \frac{1}{2} \, (w_2 - w_1) \,  g_{ij}(Q_1) \,
k \Delta x^i \, \, k \Delta x^j \,  \int^{w_2}_{w_1} \, dw  \\
   &  =  &  \frac{1}{2} \, g_{ij}(Q_1) \, \Delta x^i \, \Delta x^j
\end{eqnarray}
Similarly, the second term on the right side of
Eq.~(\ref{strainMeasure}) can be approximated as
\begin{equation} \label{SecondTerm}
\Omega(P_1,P_2) =  \frac{1}{2} \, g_{ij}(P_1) \, \Delta X^i \, \Delta X^j
\end{equation}
where the coordinates $X^i$ are the undeformed ones and $\Delta
X^i =  X^i_2 - X^i_1$. The measure of strain in
Eq.~(\ref{strainMeasure}) is then
\begin{equation}  \label{strainTensor1}
(\Delta s)^2 -(\Delta S)^2 =
   g_{ij}(Q) \, \Delta x^i \,  \, \Delta x^j \,
 - g_{ij}(P) \, \Delta X^i \,  \, \Delta X^j \,
\end{equation}
or using the flow function in Eq.~(\ref{defMapFcn}),
\begin{equation}  \label{strainTensor2}
(\Delta s)^2 -(\Delta S)^2 = \left( g_{ij}(Q) \, \frac{\partial
x^i}{\partial X^k}(P,Q) \,  \frac{\partial x^j}{\partial X^l}(P,Q)
- g_{kl}(P) \right) \, \Delta X^k \, \Delta X^l
\end{equation}
Note that $x^i$  and $X^i$ refer to to the same system of
coordinates.   I have dropped the subscripts on $Q$ and $P$ since
$Q_1$ and $Q_2$, and $P_1$ and $P_2$,  are infinitesimally close,
respectively. The quantity in parenthesis on the right side of
Eq.~(\ref{strainTensor2}) is twice the Lagrangian strain tensor:
\begin{equation}  \label{LagrangianstrainTensor}
2 \, E_{k_P l_P} = g_{ij}(Q) \, \frac{\partial x^i}{\partial
X^k}(P,Q) \,  \frac{\partial x^j}{\partial X^l}(P,Q) \, -
g_{kl}(P)
\end{equation}
From Eq.~(\ref{LagrangianstrainTensor}), it is clear that the
Lagrangian strain tensor is a two-point tensor.  Under
transformation of coordinates at point $P$,  $E_{k_P l_P}$ is a
second rank tensor, while under transformation of coordinates at
point $Q$, it is a scalar.  The deformation gradient, $\partial
x^i / \partial X^k $, is itself a two-point tensor, as can be seen
by its transformation property when coordinates at $P$ and $Q$ are
changed.

It is interesting to note that the Lagrangian strain tensor
$E_{kl}$ is conventionally thought to be a function of material
coordinates, $X^i$, which coincide with the point $P$ (in the
reference configuration)~\cite{Eringen1962,Narasimhan1992}.  The
tensor $E_{k_P l_P}$ can be taken to be a function of only the
material coordinates by using the flow mapping function in
Eq.~(\ref{defMapFcn}), which provides a mapping between all points
$P$ and their images, points $Q$, under the deformation.  I do not
pursue this interpretation below.

The first term in  Eq.~(\ref{LagrangianstrainTensor}) is the Green
deformation tensor:
\begin{equation}  \label{GreenDefTensor}
C_{kl} =  g_{ij}(Q) \, \frac{\partial x^i}{\partial X^k}(P,Q) \,
\frac{\partial x^j}{\partial X^l}(P,Q)
\end{equation}
The point $P$ is in the reference configuration at time $t=t_o$
and the point $Q$ is in the deformed state at time $t$.  In
Eq.~(\ref{GreenDefTensor}), the Green deformation tensor is
naturally a second rank tensor with respect to material coordinate
 transformations (point $P$) and it is a scalar spatial coordinate
transformations (point $Q$).  However, the Green tensor is
conventionally taken~\cite{Eringen1962,Narasimhan1992} as a
function of material coordinates by using the flow mapping
function in Eq.~(\ref{defMapFcn}).

Returning to Eq.~(\ref{strainTensor2}), and using the flow mapping
function in Eq.~(\ref{defMapFcn}), we can obtain
\begin{equation}  \label{strainTensor3}
(\Delta s)^2 -(\Delta S)^2 = \left( g_{ij}(Q)  - g_{kl}(P) \,
\frac{\partial X^k}{\partial x^i}(P,Q) \,  \frac{\partial
X^l}{\partial x^j}(P,Q) \right) \, \Delta x^i \, \Delta x^j
\end{equation}
where the Eulerian strain tensor $e_{i_Q j_Q}$ is given by
\begin{equation}  \label{EulerianStrainTensor3}
2 \, e_{i_Q j_Q} = g_{ij}(Q)  - g_{kl}(P) \, \frac{\partial
X^k}{\partial x^i}(P) \,  \frac{\partial X^l}{\partial x^j}(P)
\end{equation}
The Eulerian strain tensor $e_{i_Q j_Q}$ is  a two-point tensor
that is second rank with respect to spatial coordinates at point
$Q$ and is a scalar with respect to material coordinates at point
$P$ . Once again, however, using the flow mapping function in
Eq.~(\ref{defMapFcn}), $e_{i_Q j_Q}$ can be imagined to depend on
the material coordinates $x^i$ (point $Q$) of the deformed state.

\section{Strain Tensors in Terms of the Displacement Field}
The  Eulerian and Lagrangian strain tensors can be expressed in
terms of the displacement fields $u^{i_Q}$ and $U^{i_P}$ in
Eq.~(\ref{EulerianDispComponents}) and
(\ref{LagrangianDispComponents}).  In terms of covariant
components, the displacement field vector at $Q$ is given by
\begin{equation}\label{covariantDisplacementFiledQ}
u_{i_Q} = \Omega_{i_Q}(O,Q)- g_{i_Q j_P} \, \Omega^{j_P}(O,P)
\end{equation}
The covariant derivative of $u_{i_Q}$ with respect to point $Q$
(coordinates $x^k$) is
\begin{equation}\label{covariantDerivativeDeformationQ0}
u_{i_Q;k_Q}=\Omega_{i_Q k_Q}(O,Q)- g_{i_Q}^{~~j_P} \, \Omega_{j_P
l_P }(O,P) \, \frac{\partial X^l}{\partial x^k}
\end{equation}
where I used the chain rule for covariant differentiation since
$\Omega(O,P)=\Omega(O,P(X))$ and the material coordinates
$X^l=X^l(x^k)$ are functions of the spatial coordinates $x^k$.  It
is clear that the chain rule must be used in
Eq.~(\ref{covariantDerivativeDeformationQ0}) by considering
Cartesian coordinates.  In
Eq.~(\ref{covariantDerivativeDeformationQ0}), I also used the fact
that in Euclidean space the parallel propagator is a constant
under covariant differentiation.  The notation $\Omega_{i_Q
k_Q}(O,Q)$ means the $(i,k)$ component of the second covariant
derivative of the world function at point $Q$. In Euclidean space,
these second covariant derivatives are simply related to the
parallel propagator, see Eq.~(\ref{OmProp1})--(\ref{OmProp3}).

Using Eq.~(\ref{OmProp3}), the covariant derivative of the
displacement field in Eq.~(\ref{covariantDerivativeDeformationQ0})
becomes
\begin{eqnarray}\label{covariantDerivativeDeformationQ}
u_{i_Q;k_Q} & = & g_{i k}(Q)- g_{i_Q}^{~~ j_P} \, g_{j l}(P) \, \frac{\partial X^l}{\partial x^k}  \label{uCovariantD1} \\
            & = & g_{i k}(Q)- g_{i_Q l_P} \, \frac{\partial X^l}{\partial x^k}  \label{uCovariantD2}
\end{eqnarray}
where in the last line the metric at $P$, $g_{j l}(P)$, was used
to lower the index on the propagator. Now I multiply
Eq.~(\ref{uCovariantD2}) by the parallel propagator $g^{m_P i_Q}$,
sum on $i_Q$, and solve for the inverse displacement gradient
\begin{equation}\label{displacementGradientSolved}
\frac{\partial X^m}{\partial x^k} = g^{m_P}_{~~~ k_Q} - g^{m_P
i_Q} \, u_{i_Q ; k_Q}
\end{equation}
From Eq.~(\ref{displacementGradientSolved}), it is clear that in
Euclidean space, the deformation gradient $ \partial X^m /
\partial x^k$ is simply related to the covariant derivative of
the displacement field, $u_{i_Q ; k_Q}$.  Note however, that in a
Riemannian space, for finite deformations,  it is generally not
possible to solve for the deformation
gradient~\cite{GambiEtAl1989}.  Now, using
Eq.~(\ref{displacementGradientSolved}) in
Eq.~(\ref{strainTensor3}), I find the expression relating the
two-point Eulerian strain tensor $e_{i_Q j_Q}=e_{i_Q j_Q}(P,Q)$ to
the covariant derivatives of the  three-point displacement field
\begin{eqnarray}  \label{EulerianStrainTensor4}
(\Delta s)^2   - (\Delta S)^2 & = &  \left[
u_{i_Q ; j_Q} + u_{j_Q ; i_Q} -g^{kl}(Q) \, u_{k_Q ; i_Q} \,
u_{l_Q ; j_Q} \right] \Delta x^i \, \Delta x^j \label{strainlength} \\
  &  = & 2 \, e_{i_Q j_Q} \, \, \Delta x^i \, \Delta x^j
\end{eqnarray}
where
\begin{equation}  \label{EulerianStrainTensor5}
e_{i_Q j_Q} = \frac{1}{2} \, \left[ u_{i_Q ; j_Q} + u_{j_Q ; i_Q}
-g^{kl}(Q) \, u_{k_Q ; i_Q} \, u_{l_Q ; j_Q} \right]
\end{equation}
Equation~(\ref{EulerianStrainTensor5}) explicitly shows that the
Eulerian strain tensor $e_{i_Q j_Q}$ is a function of two points,
material coordinates at point $P$ and spatial coordinates at point
$Q$. Note that $e_{i_Q j_Q}$ is not a function of point $O$, since
by Eq.~(\ref{covariantDerivativeDeformationQ}) the covariant
derivative $u_{i_Q ; j_Q}$ does not depend on point $O$. From
Eq.~(\ref{EulerianStrainTensor5}) it is also clear that $e_{i_Q
j_Q}$ transforms as a second rank tensor under spatial coordinate
transformations and that it transforms as a scalar under material
coordinate transformations.

An analogous relation can be obtained for the Lagrangian strain
tensor by considering the covariant derivative of the displacement
field
\begin{equation}\label{covariantDerivDisplacement2}
U_{i_P ; k_P} = g_{i_P}^{~~~j_Q} \, \Omega_{j_Q l_Q}(O,Q)
\frac{\partial x^l}{\partial X^k} - \Omega_{i_P k_P}(O,P)
\end{equation}
Using the relations between the second covariant derivatives of
the world function and propagator in
Eq.~(\ref{OmProp1})--(\ref{OmProp3}), and solving for the
displacement gradient I get
\begin{equation}\label{displacementGradientSolved2}
\frac{\partial x^m}{\partial X^k} = g^{m_Q}_{~~~ k_P} - g^{m_Q  i_P} \, U_{i_P ; k_P}
\end{equation}
Inserting the displacement gradient in Eq.~(\ref{displacementGradientSolved2})
into Eq.~(\ref{strainTensor2}) gives an expression for the Lagrangian strain tensor $E_{m_P n_P} = E_{m_P n_P}(P,Q)$,
\begin{eqnarray}
(\Delta s)^2   - (\Delta S)^2 & = &  \left[
U_{m_P ; n_P} + U_{n_P ; m_P} + g^{ij}(P) \, U_{i_P ; m_P} \,
U_{j_P ; n_P} \right] \Delta X^m \, \Delta X^n \label{strainlength2} \\
  &  = & 2 \, E_{m_P n_P} \, \, \Delta X^i \, \Delta X^j
\end{eqnarray}
where
\begin{equation}\label{LagrangianStrain5}
E_{m_P n_P} = \frac{1}{2} \, \left[ U_{m_P ; n_P} + U_{n_P ; m_P}
+ g^{ij}(P) \, U_{i_P ; m_P} \, U_{j_P ; n_P} \right]
\end{equation}
Equation~(\ref{strainlength2})  shows that the Lagrangian strain
tensor $E_{m_P n_P}$ is a function of the material coordinates at
point $P$ and spatial coordinates at point $Q$. The Lagrangian
strain tensor transforms as a scalar under spatial coordinate
transformations  at point $Q$ and as a second rank tensor with
respect to material coordinate transformations at point $P$.  Note
that there are minus sign differences in
Eq.~(\ref{EulerianStrainTensor5}) and (\ref{LagrangianStrain5}).
Finally, comparing Eq.~(\ref{strainlength}) and
(\ref{strainlength2}), we have the well-known relation between the
two strain tesnors
\begin{equation}\label{tensorRelation}
E_{m_P n_P} = e_{i_Q j_Q} \, \frac{\partial x^i}{\partial X^m} \,
\frac{\partial x^j}{\partial X^n}
\end{equation}
Equation~(\ref{tensorRelation}) provides a complicated relation
between the two two-point strain tensors.   While the displacement
fields $u^{i_Q}$ and $U^{i_P}$ are related to each other by
parallel transport, see
Eq.~(\ref{EulerianDispComponents})--(\ref{LagrangianDispComponents}),
the strain tensors $E_{m_P n_P}$ and $e_{i_Q j_Q}$ are related by
two-point deformation gradient tensors, $\partial x^i /
\partial X^m$,  in Eq.~(\ref{tensorRelation}).

\section{Summary}

Conventionally, the  Eulerian and Lagrangian strain tensors are
derived either by using  shifter tensors or by using convected
(moving) coordinates. The definition of the shifter tensor makes
use of a scalar product between vectors at two different points in
space (without first parallel translating one vector to the
position of the other).  When convected coodinates are used,
vectors and tensors are associated with given coordinates in the
convected system of coordinates, rather than being associated with
a gven point in the underlying space. As discussed in the
introduction, both of these features are undesireable, when we
need to understand the transformation properties of the strain
tensors from an inertial frame to a moving frame. These
transformation properties are also needed in order to generalize
the strain tensor to Riemannian geometry for applications to
general relativity.

I have provided a derivation of the Eulerian and Lagrangian strain
tensors for finite deformations using the concepts of parallel
propagator, the world function of J. L. Synge and the three-point
displacement vector field. This derivation avoids the undesireable
features mentioned above. The derivation shows that the Eulerian
strain tensor is a two-point object that transforms as a scalar
under transformation of material coordinates and as a second rank
tensor under transformation of spatial coordinates. The derivation
also shows that the Lagrangian strain tensor behaves as a scalar
under transformation of spatial coordinates and as a second rank
tensor under transformation of material coordinates. These
transformation properties are useful in understanding how these
strain tensors transform from one frame of reference to another
moving, non-inertial frame.   The formulation presented here of
the transformation properties of these tensors is also useful for
understand the role of the reference (unstrained) configurartion
in pre-stressed materials, as discussed in the introduction.

\begin{acknowledgements}
The author thanks Dr. W. C. McCorkle, U.\ S.\ Army Aviation and
Missile Command, for suggesting investigation of the problem of
stresses in a rotating cylinder, which relied on the ideas in this
manuscript as a conceptual foundation.
\end{acknowledgements}



\begin{thebibliography}{}
\bibitem{FootNote1}I use the notation $z^k$ and $Z^k$ for the
Cartesian coordinates and also for the deformation mapping
functions.  There is no chance of confusing these two since the
context makes it clear which one is usedin a given instant.
\end{thebibliography}
\end{document}